\documentclass[aps,prl,twocolumn,showpacs]{revtex4}
\usepackage{graphicx}%
\usepackage{epsfig}
\usepackage{amsmath}%
\usepackage{amssymb}%
\usepackage{figsize}
\usepackage{subfigure}

\def\beq{\begin{equation}}

\def\eeq{\end{equation}}

\def\ksi{\xi}

\def\prl{{\sl Phys. Rev. Lett.}\ }

\def\prb{{\sl Phys. Rev. B}\ }

\begin{document}

\bibliographystyle{prsty}


\title{\Large\bf Thermal Phase Transition in Two-Dimensional Disordered Superconductors: Kosterlitz-Thouless vs Percolation}

\author{Amir Erez and Yigal Meir}

\affiliation{
 Physics Department, Ben-Gurion University, Beer Sheva 84105, Israel}
\date{\today}

\begin{abstract}
\noindent
Weakly disordered two-dimensional superconductors undergo a Kosterlitz-Thouless (KT) transition, where at a critical temperature vortices proliferate through the system and destroy the superconducting (SC) order. On the other hand, it was suggested that for large disorder the systems separates into regions of high SC order, and it is the percolation of coherence between these regions that is lost at the critical temperature. Here we demonstrate that these two descriptions are just the dual of each other. A vortex causes loss of local correlations, and thus the loss of percolation of correlations is concomitant with percolation of vortices on the dual lattice, in the perpendicular direction, i.e. the KT transition.
\end{abstract} 
\maketitle
\noindent
The interplay of disorder and superconductivity has been a subject of research from the early days of BCS theory  \cite{BCS}, when Anderson has demonstrated that weak disorder does not affect the BCS critical temperature \cite{AndersonsTheorem}. It was later argued that superconductivity indeed persists even when the single-particle states are localized  \cite{LeeMa}, but eventually, with strong enough disorder, superconductivity is destroyed. Thus the critical temperature is reduced with increasing disorder, until it is suppressed all the way to zero for large enough disorder, indicating a zero-temperature transition from a superconducting (SC) to an insulating phase. The situation is even more intriguing in two dimensions, where, even without disorder, there could be no long-range SC order at any finite temperature $T$ due to the Mermin-Wagner theorem \cite{MerminWagner}. Nevertheless, one expects a finite-temperature Kosterlitz-Thouless (KT) transition \cite{KosterlitzThouless} between a phase with power-law decaying correlations to a phase with exponentially decaying correlations. This transition is driven by the unbinding and proliferation of vortex-anti vortex pairs, which destroy the SC order. The effect of disorder on the KT transition has become even more relevant since the experimental observation of superconductor-insulator transition in thin disordered films, about two decades ago \cite{Haviland,Goldman_review}. Weak disorder should not affect the transition, according to the Harris criterion \cite{HarrisCriterion}. In the presence of strong disorder, however, it has been established theoretically \cite{Spivak,Galitzki,Ghosal,Dubi} and observed experimentally \cite{Kowal,Sacepe} that the SC order parameter fluctuates strongly across the sample, creating "SC islands", where the SC order is high, surrounded by areas of weaker SC correlations. Thus, with increasing temperature the coherence between neighboring SC islands is quenched, until percolation of coherence from one side of the sample to the other is lost, leading to
 the loss of global SC order \cite{Scalettar,DubiNature}. This description predicts that local SC order may persists even when global SC order is lost, consistent with recent experiments \cite{Gantmakher,Crane,Vales}.

 In this letter we study the thermal phase transition from a SC to a normal state in disordered two-dimensional superconductors. In particular we address the question of relevance of the KT description to this transition, and confront this description with the  percolative description. To that end we begin with the negative-U Hubbard model,

\begin{eqnarray}
\label{eq:H_Hubbard}
\mathcal{H}&=&-\sum_{\langle i,j \rangle,\sigma}t_{ij} C_{i\sigma}^\dag C_{j\sigma} + U\sum_i C_{i\uparrow}^\dag C_{i\downarrow}^\dag C_{i\downarrow}C_{i\uparrow}\nonumber\\ &+&\sum_{i,\sigma} \left( V_i - \mu \right) C_{i\sigma}^\dag C_{i\sigma} ,
\end{eqnarray}
where $\langle i,j \rangle$ indicates a sum over nearest neighbors, $C_{i\sigma}^\dag$ creates a spin-$\sigma$ electron at site $i$, $t_{ij}$ is the hopping integral, taken to be the unit of energy in the following, and $U<0$ is the on-site attractive potential. The site-specific disorder $V_i$  is taken from a uniform distribution of width $2W$ such that $V_i \in [-W,W]$, but for each realization we ensure that $\sum_i V_i = 0$, and the chemical potential $\mu$ determines the average density $n$. The choice of the Hubbard model is because it can lead, depending on parameters,  to a BCS transition, to a KT transition or to a percolation transition, and thus it is general enough not to limit a priori the possible transitions, unlike, e.g. the  XY model, which is known to give rise to a KT transition \cite{KosterlitzThouless} and may bias the results towards this particular transition.

In order to consider SC fluctuations, crucial for the description of the transition, we employ a method that  takes into account thermal phase fluctuations, but ignores the quantum ones \cite{Mayr,DubiNature}. In short, applying a Hubbard-Stratonovic transformation to the Hubbard Hamiltonian (\ref{eq:H_Hubbard}), with a local complex Hubbard-Stratonovic field, $\Delta_i$, and ignoring the temporal dependence of these fields (quantum fluctuations), the partition function becomes:

\begin{equation}
\label{eq:partition}
Z = Tr[ e^{-\beta \mathcal{H}}] = \int \mathcal{D}(\{\Delta_i,\Delta_i^*\}) Tr_f [e^{-\beta \mathcal{H}_{BdG}(\{\Delta_i\})}], \nonumber
\end{equation}

with the Bogoliubov-de Gennes Hamiltonian \cite{deGennes} $\mathcal{H}_{BdG}(\{\Delta_i\})$ given by
\begin{eqnarray}
\label{eq:H_BdG}
\mathcal{H}_{BdG} &=& -\sum_{\langle i,j \rangle,\sigma}t_{ij} C_{i\sigma}^\dag C_{j\sigma} +\sum_{i\sigma} \left( V_i - \mu \right) C_{i\sigma}^\dag C_{i\sigma} \nonumber \\
 &+&  \sum_{i\sigma} \left( U_i C_{i\sigma}^\dag C_{i\sigma} + \Delta_i C_{i\uparrow}^\dag C_{i\downarrow}^\dag + \Delta_i^*C_{i\downarrow}C_{i\uparrow} \right).\nonumber
\end{eqnarray}

Here $Tr_f$ traces the fermionic degrees of freedom over the single-body Hamiltonian $\mathcal{H}_{BdG}$ and can be evaluated exactly using its eigenvalues \cite{Weisse}. The integral over the fields $\{\Delta_i,\Delta_i^*\}$ can then be calculated using the (classical) Metropolis Monte-Carlo (MC) technique \cite{Metropolis}. One should note that unlike the usual BdG approach, here $\Delta_i$ are auxiliary fields, and except at zero temperature where the saddle point evaluation of the partition function gives rise to the BdG solution, they are generally different from the local SC order parameter $<C_{i\downarrow}C_{i\uparrow}>$.

This procedure allows calculation of any ensemble-averaged quantity. We first calculate  the average correlation function $D_{ij}\equiv<\cos(\theta_i-\theta_j)>$, where $\theta_i$ is the phase of the SC order parameter $<C_{i\downarrow}C_{i\uparrow}>$, and $i$ and $j$ are on the opposite edges of the sample. Bulk SC order exist only if these phases are correlated, hence the decay of $D_{ij}$ to zero signals the loss of bulk SC order in the system. This calculation was carried out in two ways. $D_{ij}$ can be calculated directly using the MC procedure.  Alternatively, we first calculate the whole distribution of $\theta_i-\theta_j$, for any pair $i$ and $j$. We then compare this distribution, for every temperature, to the one expected from a single Josephson junction,
\begin{equation}
\label{eq:XY_Hamiltonian}
\mathcal{H}_{ij}^{eff}=-J_{ij}(T) \cos(\theta_i-\theta_j).
\end{equation}

\begin{figure}[tbp]
\includegraphics[width=7truecm]{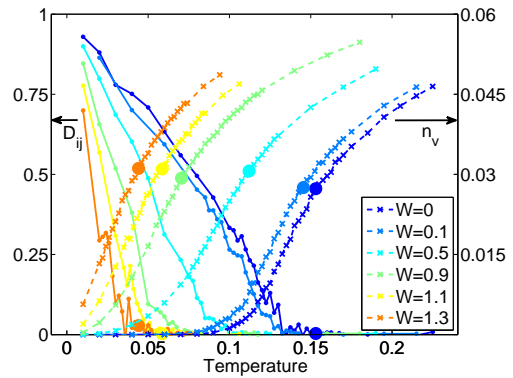}
\caption{\small The edge-to-edge phase correlation $D_{ij}$ (left axis) and the vortex density $n_v$ (right axis) as a function of temperature, for different values of disorder for $18\times 18$ system, and $U=-4$, and values of disorder depicted in the legend. With increasing disorder $T_c$, where bulk coherence is lost and vortices start to proliferate, shifts to lower values, but the curves still look similar. The circles denote the percolation temperature $T_p$ (see text).
\label{fig:cos}}
\end{figure}

The excellent fit to the data (not shown) allows us to determine the (temperature dependent) effective Josephson coupling $J_{ij}(T)$. Once this parameter is determined, $D_{ij}$ can be directly evaluated, using the Hamiltonian (\ref{eq:XY_Hamiltonian}) to be $I_1(J_{ij}/T)/I_0(J_{ij}/T)$, where $I_n(x)$ is Bessel function of the first kind. Both approaches give almost exactly the same results,  plotted in Fig.~\ref{fig:cos}. As can be seen in the figure, increasing disorder suppresses the critical temperature $T_c$, as expected. Except for the change in $T_c$, the decay of correlations at finite disorder looks very similar to the zero disorder case, expected to be described by the KT transition.

In order to quantify this observation further, we calculate the sample-averaged vortex (and anti-vortex) density, $n_v$. The local vortex density for a single snapshot during the MC procedure is determined by the clockwise integration of the phase of the SC order parameter around a single plaquette. The value of this integral can assume only integer multiples of $2\pi$, indicating the existence of vortices $(n>0)$, anti-vortices $(n<0)$, or no vorticity $(n=0)$. In Fig.~\ref{fig:cos} we also plot $n_v$ as a function of temperature. As can clearly seen, the loss of global phase coherence occurs when vortices start to proliferate, for {\sl all values of disorder}, indicating a KT-type transition. 

In order to further identify the KT characteristics of the transition, we adopt a particularly elegant analysis proposed by Polkovnikov, Altman and Demler \cite{PAD_exponent}, and used in experimental analysis by Dalibard et al. \cite{Dalibard_uses_PAD}. The idea is to perform the double sum over all pairwise correlations $D_{ij}$, up to the system size $L$,

\begin{equation}
A(L,T) \equiv \frac{1}{L^2} \sum_{i=1}^{L} \sum_{j=1}^{L} D_{ij}(T) \sim L^{-2\alpha(T)},
\end{equation}
where the second relation defines the exponent $\alpha$.
As $T\rightarrow 0$ we expect $D_{ij}\rightarrow 1$ meaning that $A(L,T\rightarrow 0)=1$ and therefore $\alpha(T\rightarrow 0)=0$.
In the opposite limit, when $T > T_c$ then $D_{ij}$ decays exponentially with the distance $|i-j|$, leading to $A(L,T\ge T_c)\rightarrow L^{-1}$ and $\alpha\rightarrow 0.5$.
At the KT transition (see \cite{PAD_exponent}), a universal jump from $\alpha=0.25$ to $\alpha=0.5$ is expected.

\begin{figure}[!h]
 \centering
\includegraphics[width=6cm]{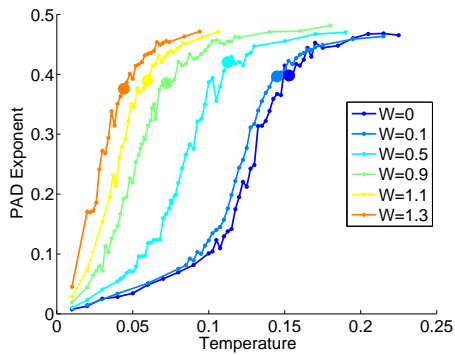}
\caption{The PAD exponent \cite{PAD_exponent} $\alpha$ for $U=-4$, $18\times 18$ system for different values of disorder, exhibiting an abrupt change, signalling the transition. Large dots correspond to $T_p$ from the percolation picture (see text). }
  \label{fig:PAD}
\end{figure}

Fig.~\ref{fig:PAD} depicts our results for the PAD exponent $\alpha$ as function of temperature, for different values of disorder. All the curves (except for the highest disorder, probably already on the insulating side) exhibit a very similar behavior -- a moderate rise from $\alpha=0$ to $\alpha\simeq0.1$ and then a rather abrupt rise towards $\alpha\simeq0.5$, demonstrating a change from power-law decay of correlations to exponentially decaying correlations, again consistent with the KT description. We attribute the deviation of the value of the exponent where the abrupt change occur from the expected value ($\alpha=0.25$) to the finite size of our system, as the coherence length $\xi_0$, which is approximately given by $\hbar v_F/\Delta$, where $v_F$ is the Fermi energy and $\Delta$ the SC gap, is not much smaller than the system size (about a factor of 2 or 3). The fact that the same behavior is observed at zero disorder, where the KT description is indeed expected to hold, supports this explanation.

\begin{figure}[!ht]
 \centering
  \subfigure[$\ T=0.02$, deep in the SC state]{\includegraphics[clip,width=0.48\hsize]{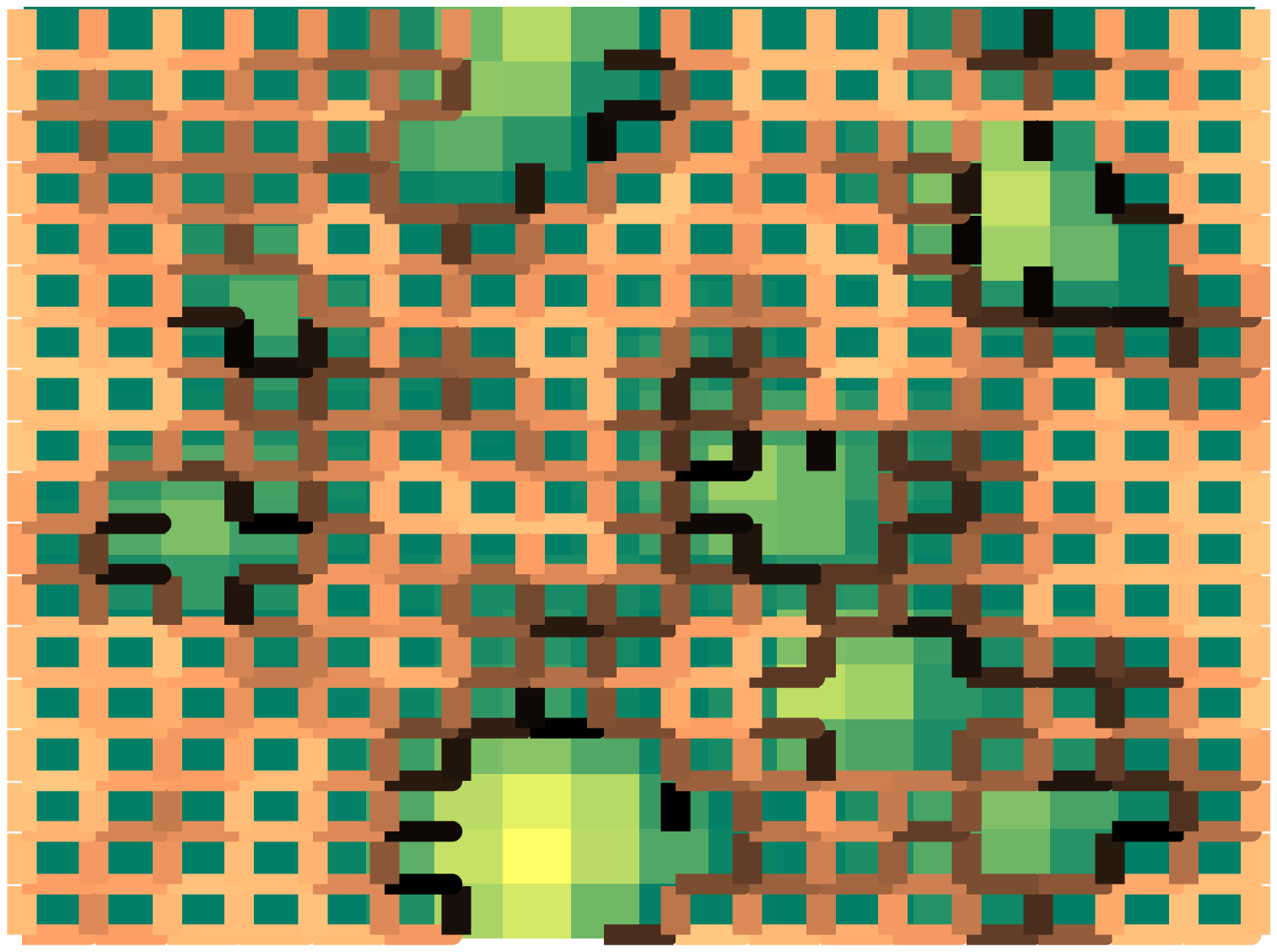}}
  \subfigure[$\ T=0.048$, in the SC state close to the transition]{\includegraphics[clip,width=0.48\hsize]{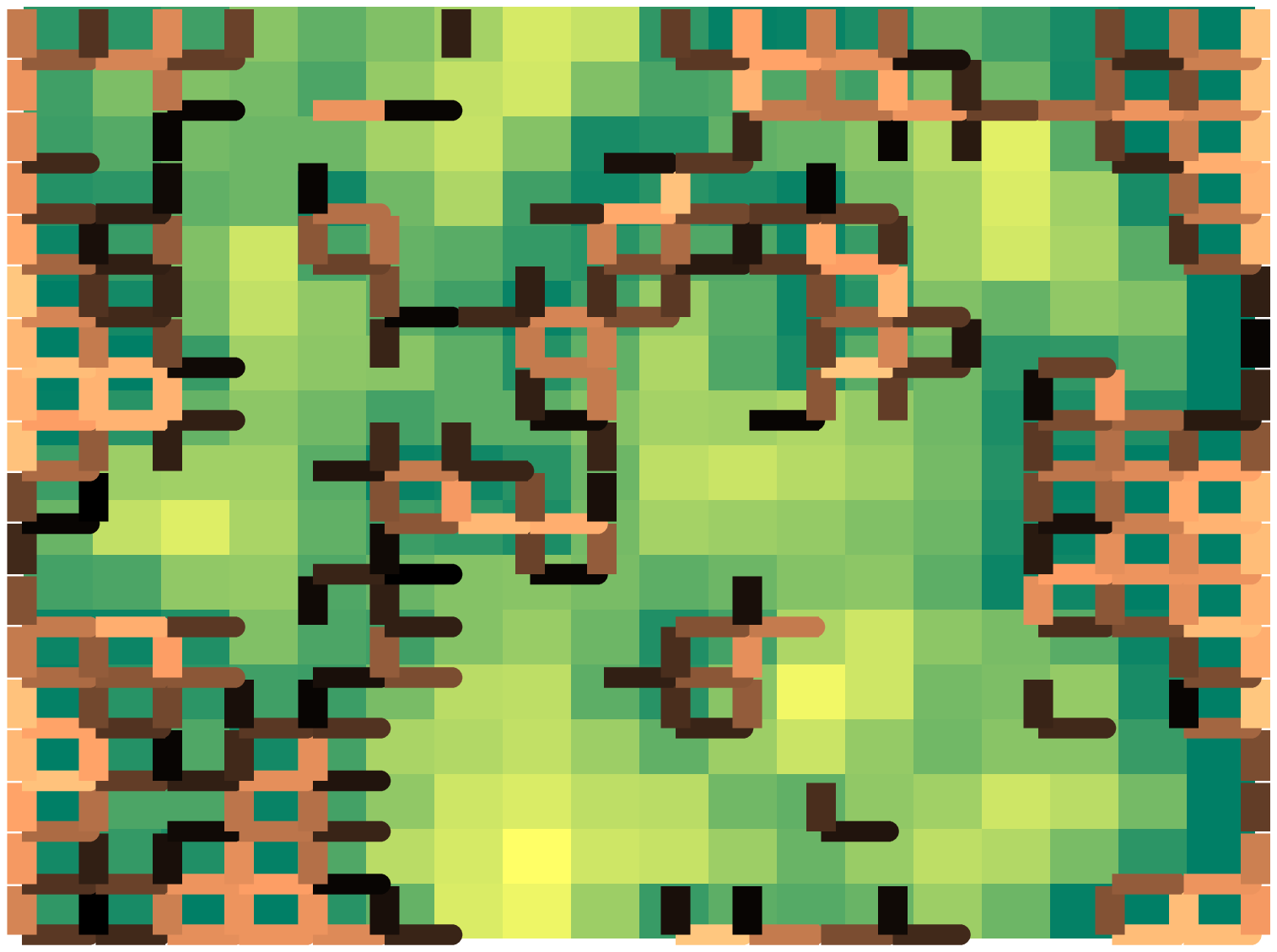}}
  \subfigure[$\ T=0.052$ insulating state, just crossed the transition]{\includegraphics[clip,width=0.48\hsize]{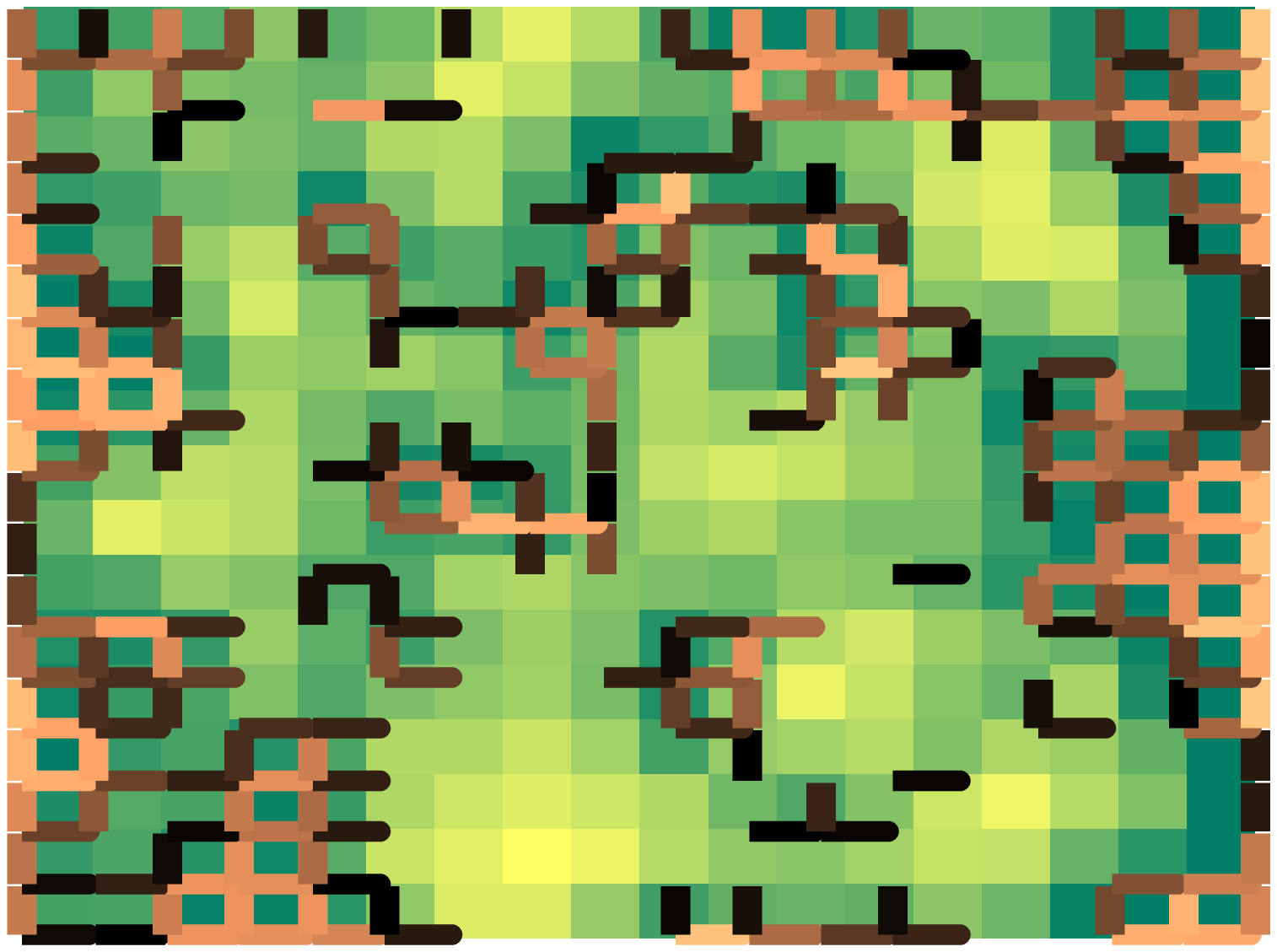}}
  \subfigure[$\ T=0.074$ deep in the insulating state]{\includegraphics[clip,width=0.48\hsize]{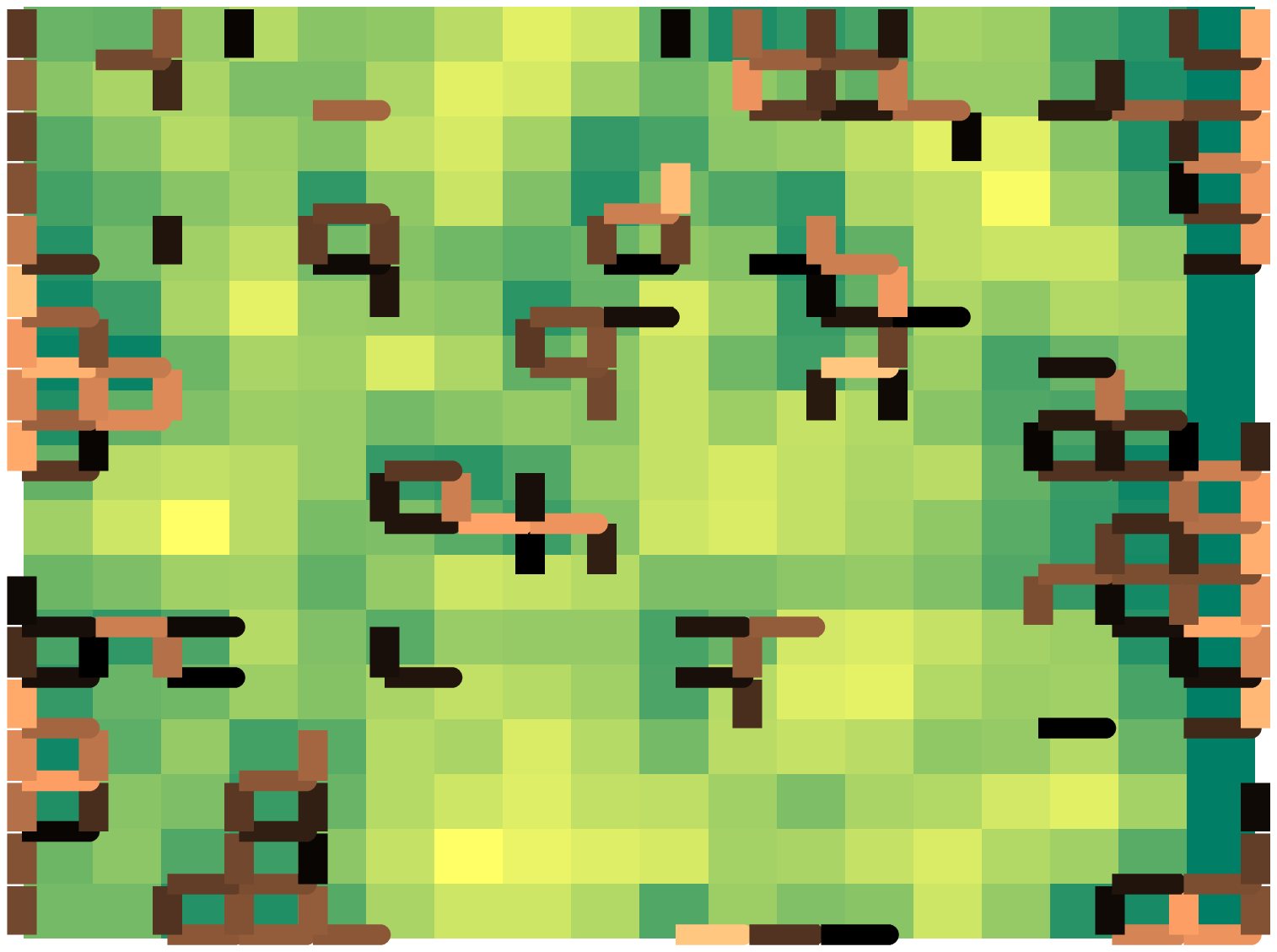}}
  \caption{ Percolation transition for a $18 \times 18$ system ($U=-4$) at disorder $W=1.3$: Nearest neighbor links $D_{ij}$ (lines) colored from orange for strong correlations, to black for weaker correlations (close to cutoff $d_{ij}=0.45$), and disconnected for correlations below the threshold. Temperature changes from deep in the SC regime (a) to deep in the insulating regime (d). Local vortex density $n_v$ (background) colored from dark green for low density, to bright yellow for high density, demonstrating the correlation between the local $D_{ij}$ and the local $n_v$ (see also Fig.~\ref{fig:Jnv}).}
  \label{fig:perc}
\end{figure}

Having established the relevance of the KT description for finite disorder, we now turn to check whether the same data can be described by the percolation picture. According to the discussion above, we can determine the effective Josephson coupling $J_{ij}$ between nearest neighbors. With increasing temperature $J_{ij}$ decays, until correlations are lost, around $J_{ij}(T)\simeq T$ (corresponding to $D_{ij} \simeq 0.45$). Thus we can set this value as a threshold for an existing coherence link between nearest-neighbors, and determine the temperature where edge-to-edge percolation is lost, $T_p$. Fig.~\ref{fig:perc} demonstrates the connectivity of the network for temperatures from below to above the percolation temperature, where the existing links are colored by strength (of $J_{ij}$ or $D_{ij}$), and the missing links are those with $D_{ij}$ below the above threshold. The percolation temperature $T_p$ can be determined for each disorder value, and, in fact, corresponds to an average link density  of $1/2$, consistent with bond percolation on a square lattice. $T_p$ is also depicted as circles in Fig.\ref{fig:cos}, demonstrating that it is indeed in the region where bulk correlations decay and vortices start to proliferate.


\begin{figure}[!ht]
 \centering
  \subfigure{\includegraphics[clip,width=0.49\hsize]{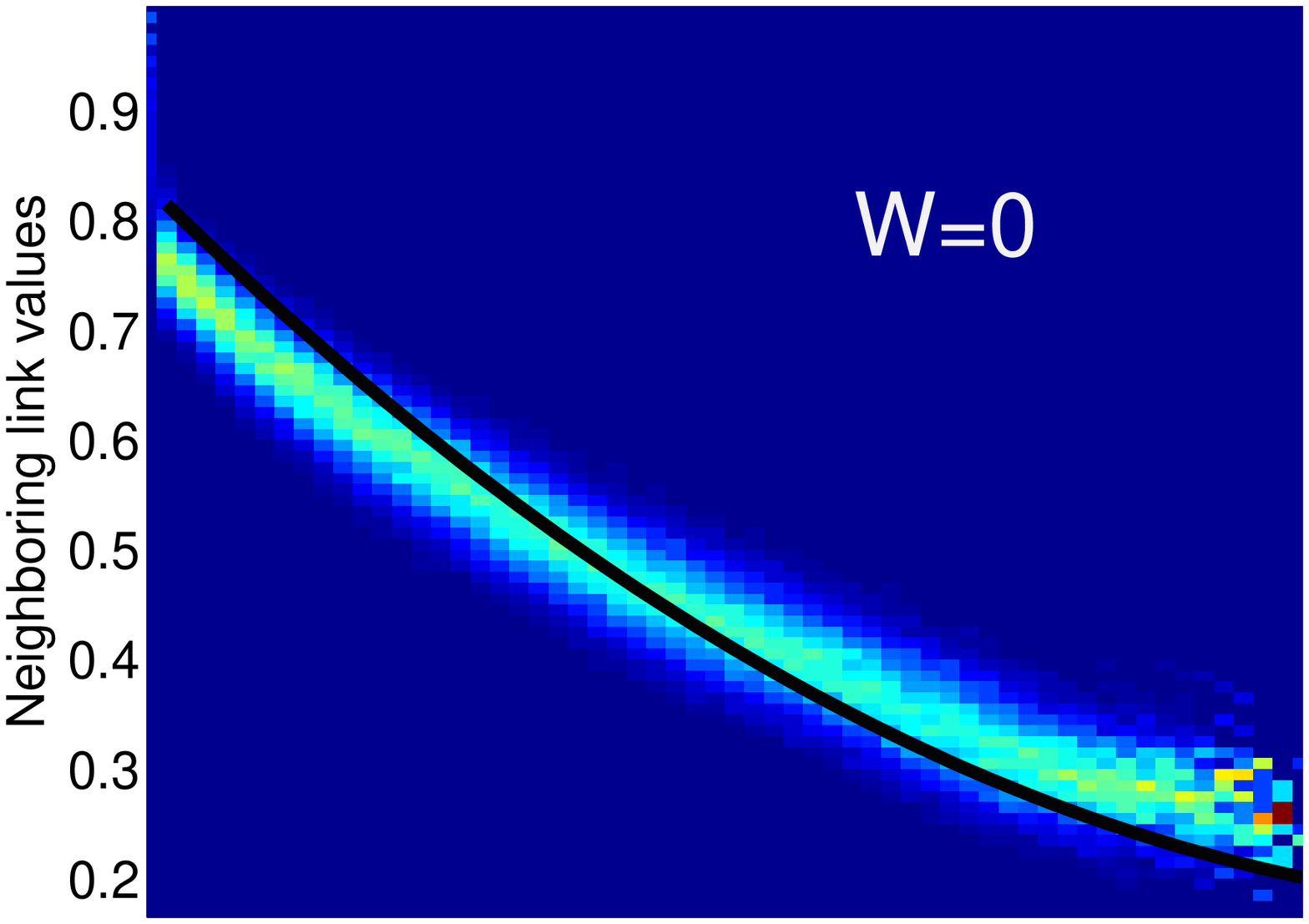}}
  \subfigure{\includegraphics[clip,width=0.49\hsize]{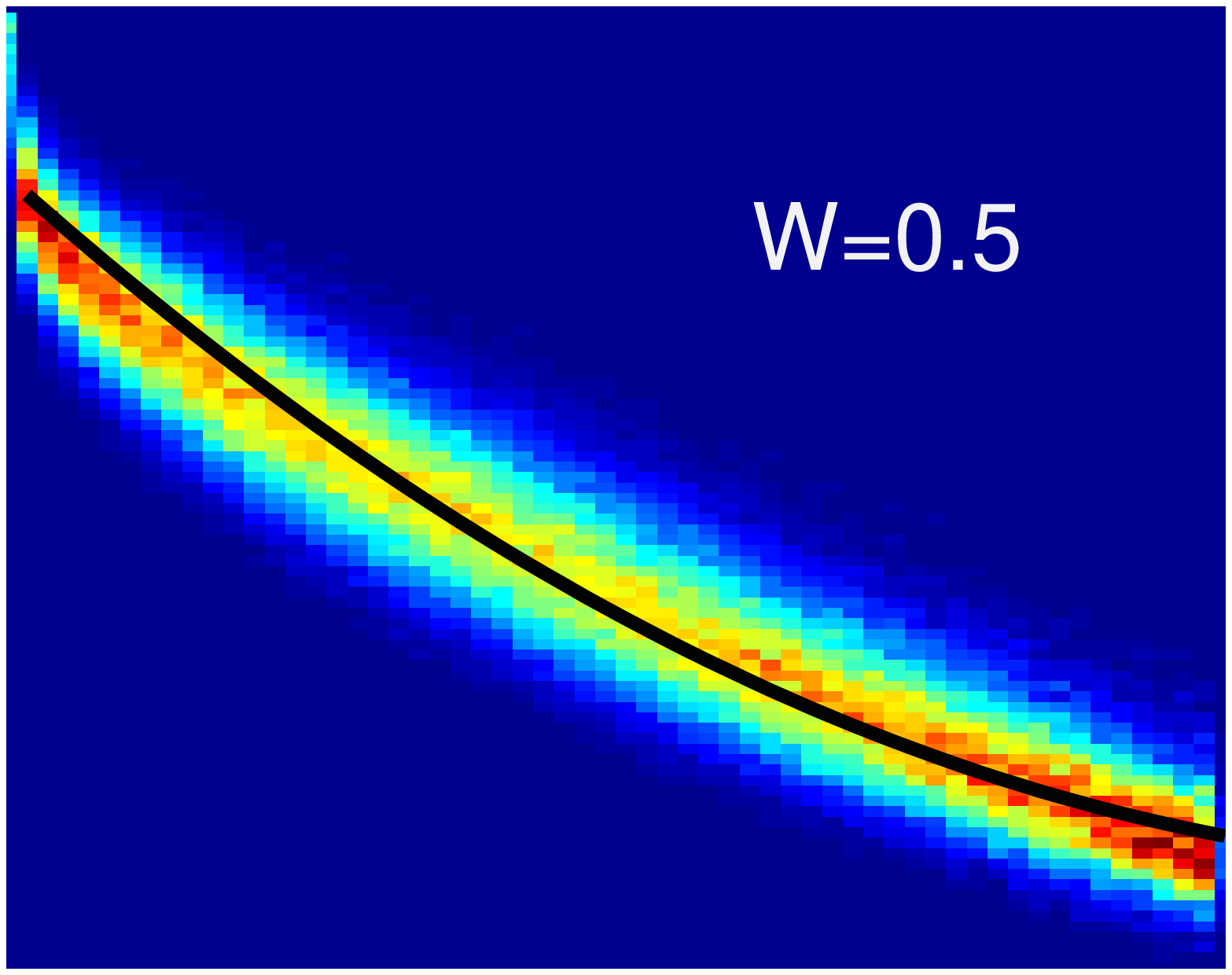}}
  \subfigure{\includegraphics[clip,width=0.49\hsize]{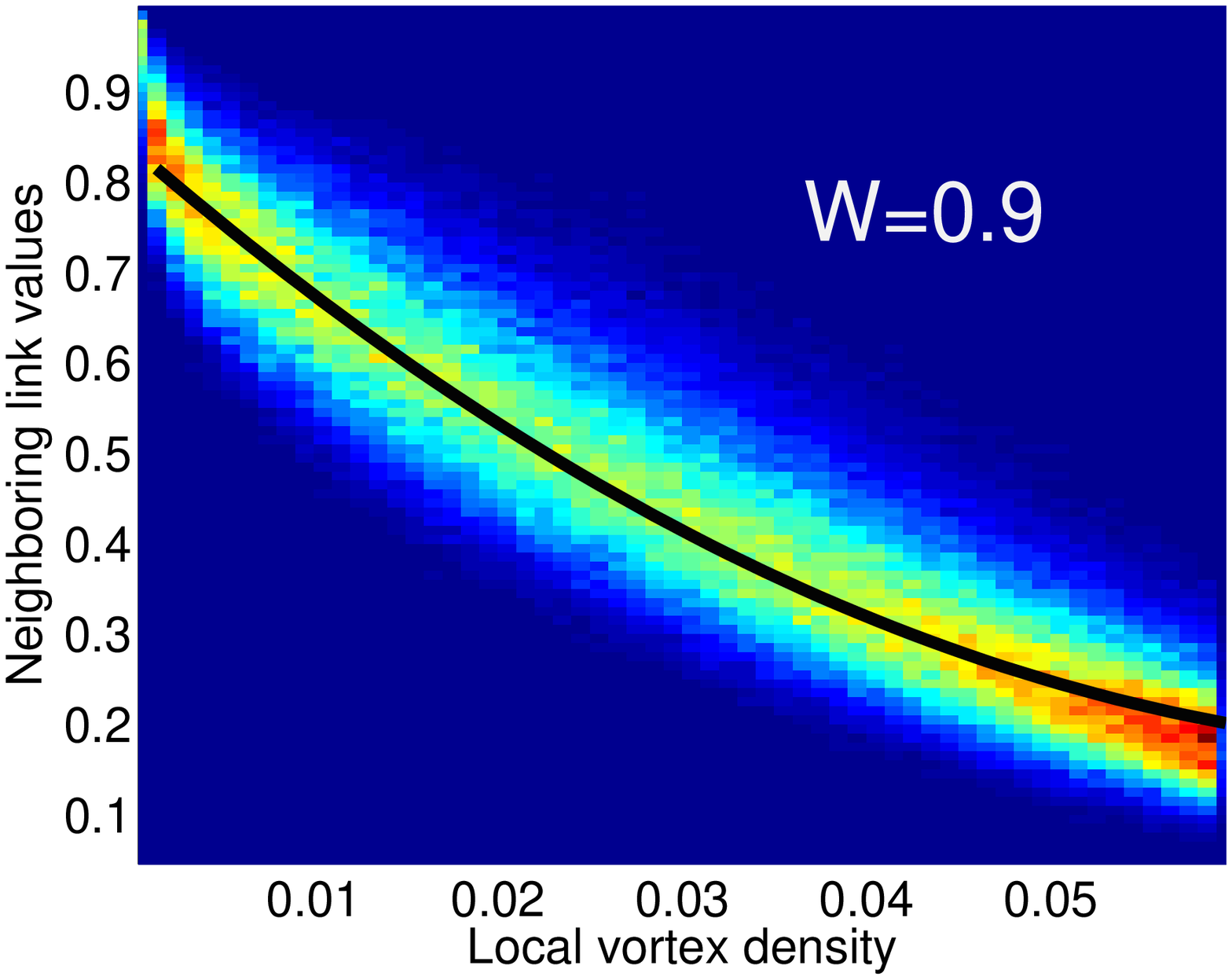}}
  \subfigure{\includegraphics[clip,width=0.49\hsize]{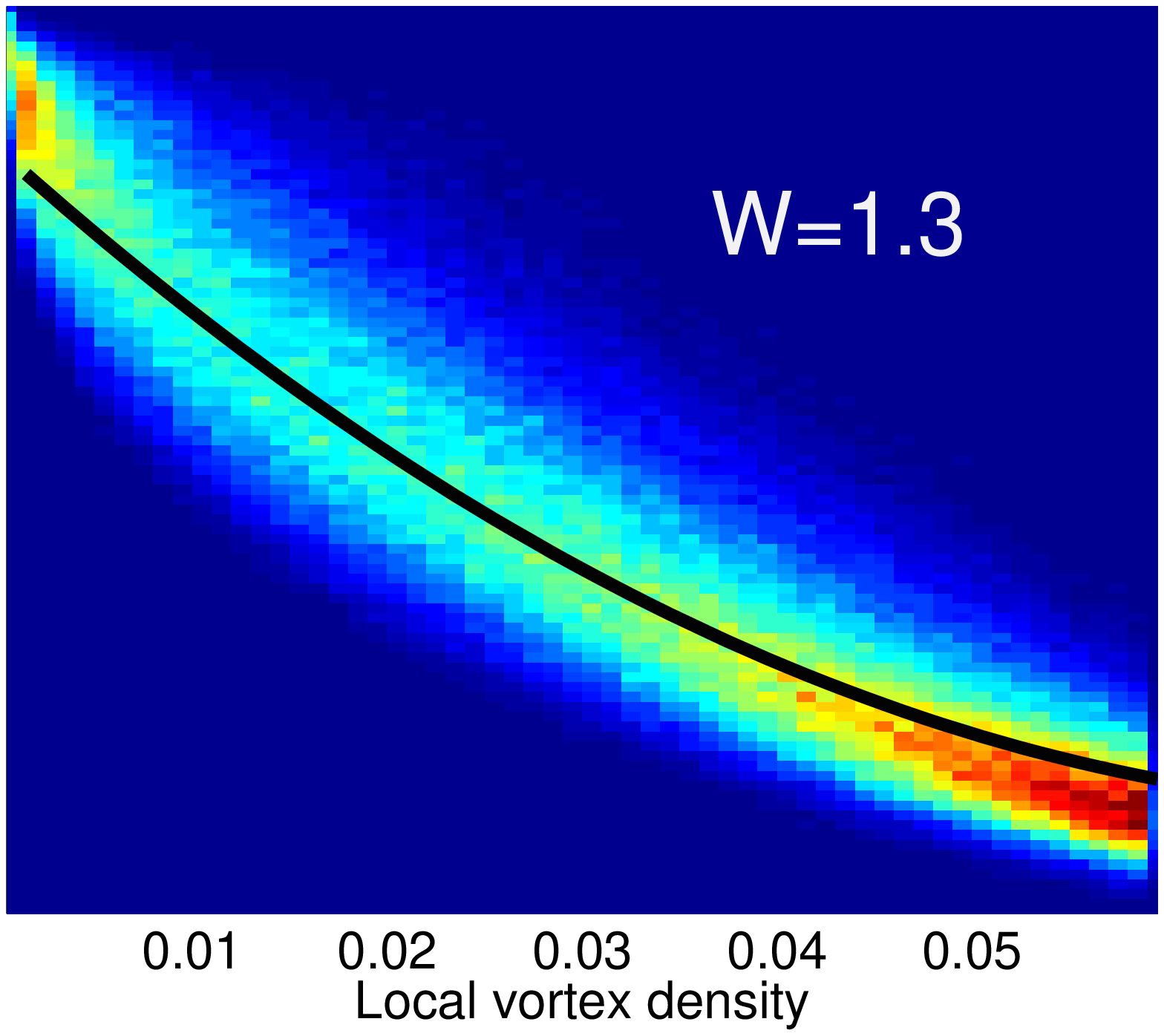}}
  \caption{The joint distribution of the local SC correlations and the local vortex density (normalized to unity at each vortex density). The strong correlation between the two values persists with increasing disorder, with the same functional form (solid line).}
  \label{fig:Jnv}
\end{figure}

In order to quantify this last statement further, we have also plotted in Fig.~\ref{fig:perc} the local vortex density $n_v$, in brighter shades of green for higher density. As can be clearly seen, as the local $n_v$ increases, the neighboring $D_{ij}$ decrease and disconnect. In Fig.~\ref{fig:Jnv} we plot the dependence of the local $J_{ij}$ on the local $n_v$, for different values of disorder. This dependence is observed to be independent of disorder (except for small corrections at zero disorder and a slightly wider distribution with increasing disorder), leading to the local $J_{ij}$, up to some small fluctuations, being a unique function of the local vortex density. This observation allows us to relate the two pictures: in order for the percolation path of the $J_{ij}$'s to disconnect, a perpendicular line connecting points of  $n_v$, larger than some threshold value, has to be formed. Thus loss of percolation of the $J_{ij}$, corresponds, in fact, to percolation of the vortices in the dual lattice in the perpendicular direction. Since this necessitates a particular density of vortices, it can only occur when vortices proliferate, i.e. at the KT transition. This point is further verified by marking the percolation temperature on the curves for the PAD exponent, Fig.~\ref{fig:PAD}. For all values of disorder, this temperature corresponds to $\alpha\simeq0.38$, midway between the values $0.25$ and $0.5$, exactly where the KT transition should occur.

\begin{figure}[!ht]
 \centering
  \includegraphics[width=6cm]{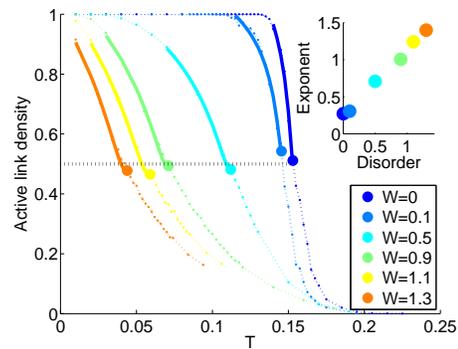}
  \caption{Link density $p$ as a function of temperature $T$. Large dots indicate $T_p$ and agree well with $p = 0.5$ as expected from percolation theory. Solid lines are  fits according to equation ~\ref{eq:pvsT}. Inset: exponents ($\tilde{B}$ deduced from the fit).}
  \label{fig:n_AL}
\end{figure}

If these two transitions describe the same physics, then a curious question arises. For the KT transition the critical divergence of the SC correlation length $\ksi$ is expected to be $\ksi(T) \sim  e^{B /\sqrt{T_c/(T_c-T)}}$ where $T_c$ the critical temperature and $B$ is some non-universal number. On the other hand, near the percolation transition one expects $\ksi(p) \sim  (p_c-p)^{-\nu}$ where $p$ is the link density (i.e. the concentration of links for which $D_{ij}$ is above the threshold), $p_c$ its critical value, and $\nu$ the correlation length critical exponent. Identifying the two transition, i.e. the critical temperature $T_c$ with the percolation temperature $T_p$, and the two critical behaviors, implies a very specific dependence of the link density on temperature
\begin{equation}
p(T) = p_c + A e^{-\tilde{B} \sqrt(\frac{T_p}{T_p-T})}.
\label{eq:pvsT}
\end{equation}
The dependence of $p$ on $T$ is displayed in Fig.~\ref{fig:n_AL}, along with the solid lines, given by Eq.~\ref{eq:pvsT}, with the resulting values of $\tilde{B}$  in the inset. The excellent agreement between the observed link density and the result of Eq.~\ref{eq:pvsT} gives further credence to the intimate connection between the KT transition and the percolation transition. This peculiar dependence of the link density on temperature indicates that as the system approaches the KT transition, the local values of $J_{ij}$ decrease rapidly, as the number of free vortices increases.  Alternatively, when temperature is reduced through the critical temperature there is an avalanche of correlations through the system. Such behavior cannot be captured, for example, by a disordered  X-Y model, where the values of the local Josephson couplings are fixed and do not change.

It would be interesting to investigate the critical behavior as one increases the thickness of the sample. Here one may expect a crossover from the KT behavior to the mean-field BCS description. Whether percolation still plays a role and how it is related to the BCS transition is a question that we plan to explore in the future.

\par This research has been funded by the ISF.

\end{document}